\documentclass[
 preprint,
 amsmath,amssymb,
 aps
]{revtex4-2}

\usepackage{graphicx}
\usepackage{dcolumn}
\usepackage{bm}
\usepackage{hyperref}
\usepackage{booktabs}

\begin{document}

\title{Artificial Intelligence and Market Entrant Game Developers}

\author{Seonbin Jo}
\affiliation{Pohang University of Science and Technology, Pohang, Republic of Korea}

\author{Woo-Sung Jung}
\affiliation{Pohang University of Science and Technology, Pohang, Republic of Korea}

\author{Jisung Yoon}
\email{jsyoon@kdischool.ac.kr}
\affiliation{KDI School of Public Policy and Management, Sejong, Republic of Korea}

\author{Hyunuk Kim}
\email{hkim6@elon.edu}
\affiliation{Elon University, Elon, NC, USA}

\begin{abstract}

Artificial Intelligence (AI) is increasingly being used for generating digital assets, such as programming codes and images. Games composed of various digital assets are thus expected to be influenced significantly by AI. Leveraging public data and AI disclosure statements of games, this paper shows that relatively more independent developers entered the market when generative AI became more publicly accessible, but their purposes of using AI are similar with non-independent developers. Game features associated with AI hint nuanced impacts of AI on independent developers. 

\end{abstract}



\maketitle

\section{Introduction}

Artificial Intelligence (AI) has been reshaping our society. Prior evidence indicates that its impacts are heterogeneous by skill, occupation, and demographic group~\cite{bick2024rapid,brynjolfsson2025canaries}. Within creative domains such as art and music, AI can generate outputs comparable to human works~\cite{doshi2024generative,zhou2024generative,haase2023artificial}, resulting in a lower entry barrier and the democratization of content production~\cite{tang2024exploring}. In this paper, we report interesting observations on games and their developers to show how AI has transformed an existing creative market. Game development requires diverse expertise in creative domains and has been conducted not only by big firms taking advantage of their large employee pool and budget but also by talented individuals with fewer resources who are called independent ({\itshape indie}) developers~\cite{martin2009independent}. 

Indie developers mostly form a small team or work individually and are thus likely to have different attitudes toward new technologies, as supported by a recent qualitative study on online posts~\cite{panchanadikar2024m} and the Technology–Organization–Environment (TOE) framework~\cite{tornatzky1990process}, especially its organizational dimension including size and financial resources. Large organizations may be able to better implement innovations~\cite{damanpour1992organizational} but frequently face structural inertia delaying decision making processes~\cite{hannan1984structural}. Entrepreneurial ventures, similar to indie developers, also listed financial resources as a factor of AI adoption~\cite{zielonka2024exploring}. Motivated by the literature, this paper focuses on recent entrant game developers who would be affected by AI and poses the following questions:

\begin{itemize}
\item {RQ1}: Did relatively more indie developers enter the game market than non-indie developers after generative AI became more accessible to the public?
\item {RQ2}: Did indie and non-indie game developers use AI for different creative purposes?
\item {RQ3}: Did indie and non-indie game developers use AI for different features?
\end{itemize}

\section{Data and Methods}
We collected the metadata of games created by a single developer in Steam, the world's largest gaming platform. The metadata contains published times, descriptions, and tags that correspond to game features. To examine the impact of AI on market entrants, only the first game of each developer was considered for analyses. The type of a developer (`indie' or `non-indie') is determined whether the first game has the `indie' tag. We identified 28,500 indie and 22,045 non-indie developers whose first game was published between January 2018 and June 2025. 

Steam's new AI policy, which was imposed in January 2024, asks all developers to report how AI is used to generate game content so that allows us to compare actual AI usage of indie and non-indie developers. We set ten categories -- Image, Music, Voice, Code, Text, Writing, Character, Gameplay, Animation, and Personalization -- and labeled each AI disclosure statement by usage. A game can have multiple AI usage labels. 

To answer RQ1, we used the numbers of market entrant indie and non-indie developers before 2024 for estimating the parameters of the CausalImpact model~\cite{brodersen2015inferring}, a Bayesian structural time series approach, and constructed two counterfactual time series from January 2024 to June 2025 with the estimated parameters. If the impact of AI is negligible, the observed and counterfactual series should closely align in a time series plot. We additionally set the number of Steam users (\url{https://steamdb.info/}) as a control time series because it can serve as a proxy for the overall platform activity and is significantly associated with the numbers of market entrant indie and non-indie developers between the years of 2018 and 2023 (Ordinary Least Squares regression; Coefficient p-value < 0.01). Yearly seasonality was also considered. 

To answer RQ2 and RQ3, we implemented the chi-square test of homogeneity for AI usage labels and calculated odds ratios from 2$\times$2 contingency matrices of which variables are developer type (indie or non-indie) and AI (used or not) for each tag. An odds ratios largely deviating from 1 indicates significantly different AI usage between indie and non-indie developers. 

\begin{figure*}[h]
  \centering
  \includegraphics[width=\linewidth]{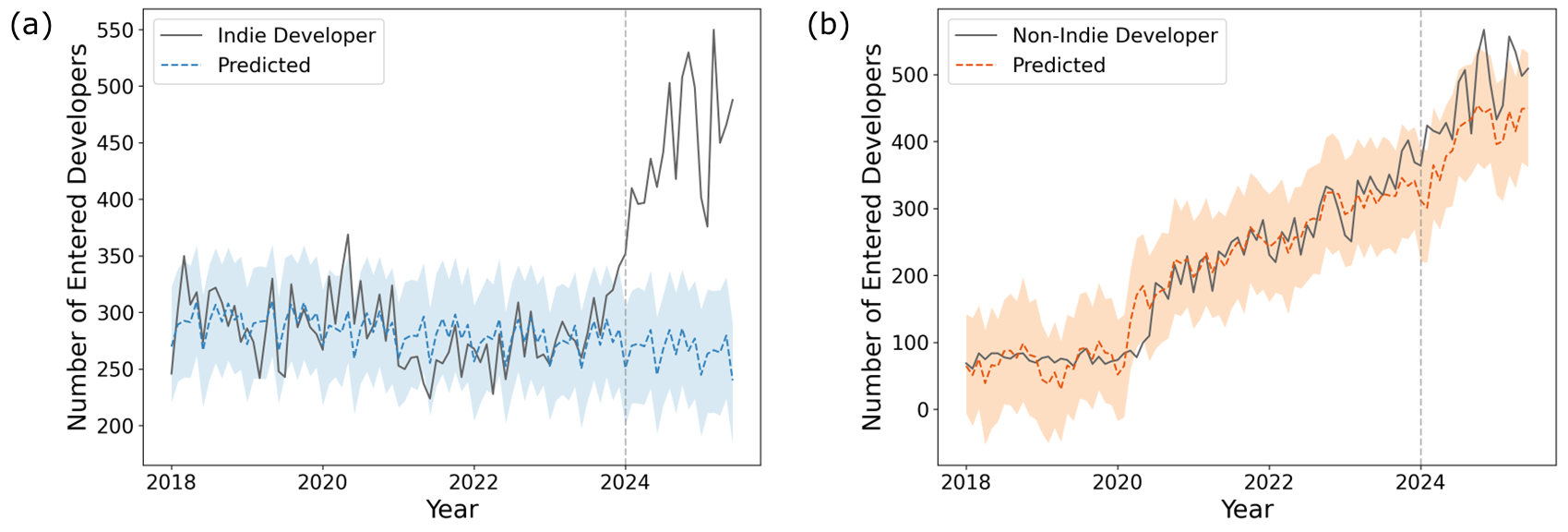}
  \caption{The observed numbers and counterfactual predictions of entrant (a) indie and (b) non-indie developers. Gray dashed lines indicate the first week of 2024, and the colored areas show 95\% confidence intervals.}
\end{figure*}

\section{Results}

A substantial increase of the number of newly entered indie developers is observed around January 2024 when Steam acknowledged extensive AI usage in game development and introduced the disclosure policy (Fig. 1a). This increase for indie developers is not explained by the previous trend and predictions, whereas the number of newly entered non-indie developers is mostly within the predicted range (Fig. 1b), suggesting that relatively more indie developers entered the market in 2024 and the first half of 2025. 

\begin{figure}[h]
  \centering
  \includegraphics[width=0.6\linewidth]{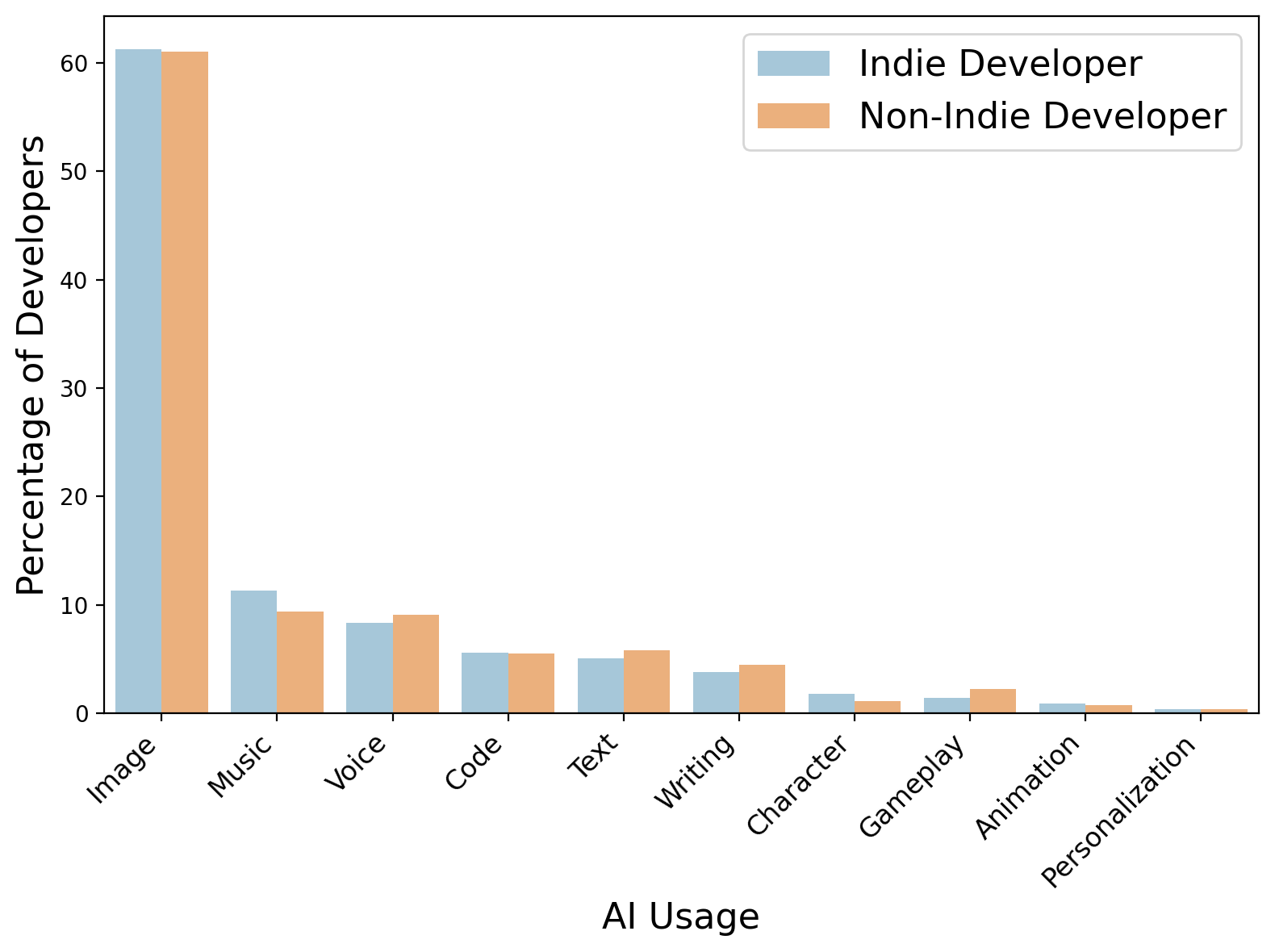}
  \caption{The distributions of AI usages in the selected Steam games published between January 2024 and June 2025.}
\end{figure}

Then, what elements did indie developers use AI for? Did AI facilitate code development most? We found that the prevalent purpose is to generate image followed by music, voice, and code (Fig. 2), and non-indie developers also used AI for similar creative purposes (Chi-square test of homogeneity; p-value = 0.48). Motivated indie developers might use AI to learn programming quickly and generate elements to realize their ideas, leading to more market entrants but a high AI usage for creating image not for code. 

\begin{figure}[h]
  \centering
  \includegraphics[width=0.6\linewidth]{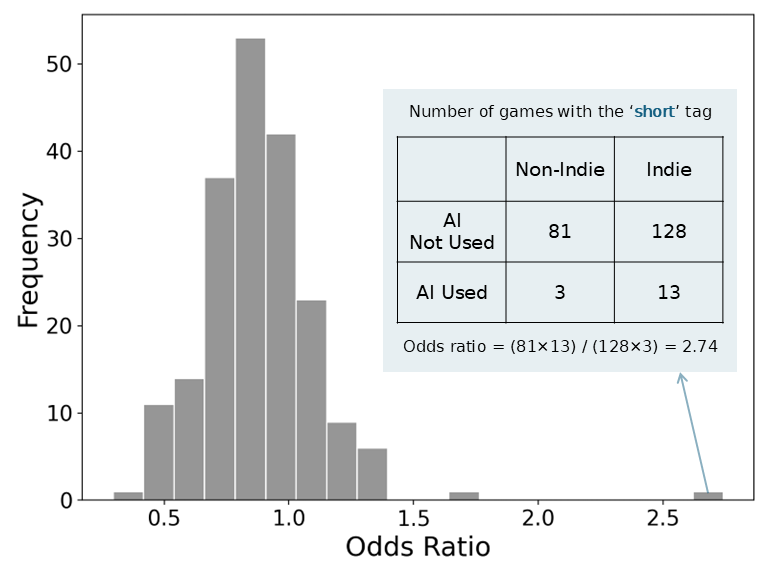}
  \caption{The distribution of the odds ratios calculated from 2$\times$2 contingency matrices for game counts by developer type and AI usage.}
\end{figure}

Although indie and non-indie developers use AI for similar creative purposes, features that they incorporate into games may vary. We quantified the association between developer type and AI usage by calculating the odds ratio for each tag. The pooled odds ratio of 198 tags appearing in more than 100 distinct indie developers’ games is 0.85 (Cochran-Mantel-Haenszel test; p-value < 0.01), suggesting that game features are associated with AI differently, especially less by indie developers on average. The calculated odds ratios for the selected tags are distributed well around the pooled value (Fig. 3), while they do not follow a normal distribution (The null hypothesis of a normality test is rejected with a p-value smaller than 0.01). The distribution of the calculated odds ratios indicate that some tags are much more or less likely to be observed in indie games. The `short' tag (i.e., short in game play length) has a substantially high odds ratio of 2.74, while tags about adult content (e.g., nsfw, mature) return a low odds ratio. The tags of the 10 highest and lowest odds ratios are listed in Table 1.  

\begin{table}[!]
  \caption{The game tags of the 10 highest and lowest odds ratios.}
  \footnotesize
  \begin{tabular}{ccc}
    \toprule
    Tag & Total Game Count & Odds Ratio\\
    \midrule
    short & 225 & 2.74\\
    beat 'em up & 331 & 1.73\\
    2.5d & 589 & 1.39\\
    racing & 554 & 1.38\\
    runner & 460 & 1.38\\
    aliens & 486 & 1.33\\
    flight & 267 & 1.28\\
    deckbuilding & 538 & 1.28\\
    medieval & 1017 & 1.22\\
    parkour & 399 & 1.21\\
    \midrule
    crafting & 804 & 0.52\\
    nonlinear & 440 & 0.51\\
    isometric & 508 & 0.51\\
    comic book & 363 & 0.50\\
    fast-paced & 200 & 0.49\\
    psychedelic & 373 & 0.48\\
    lgbtq+ & 360 & 0.48\\
    nsfw & 526 & 0.46\\
    mature & 488 & 0.46\\
    level editor & 280 & 0.29\\
  \bottomrule
\end{tabular}
\end{table}

\section{Conclusion}
Our study shows that relatively more indie developers entered the market along with generative AI (RQ1) while they used AI for similar creative purposes with non-indie developers (RQ2). However, the extent to which a tag is associated with AI in their games varies widely (RQ3).
Even though the stated uses of AI are similar between indie and non-indie developers (RQ2), AI likely lowers entry barriers by substituting for asset-heavy work—especially image creation—thereby benefitting resource-constrained indie developers more. 

Consistent with this potential explanation, the strong positive association between AI and the ‘short’ tag in indie games may add more context to RQ1 in a way that indie developers could learn game development and focus on implementing their ideas quickly. Another interesting finding is the negative association between AI and tags about adult content in indie games. Of course, tags cannot fully represent games. Even though some games have the same tags, they are surely different creative products. In addition, tags are not solely independent. For example, tags about adult content often co-occur in games. Analyzing further metadata (e.g., textual descriptions) and clusters of correlated tags would lead to better comparisons.  

Despite these limitations, our letter demonstrates positive aspects of AI on individuals motivated to develop creative digital products, as opposed to concerns on the abundance of harmful content online~\cite{wei2024exploring}. Conducting a survey for market entrant indie and non-indie developers will provide deeper insights into the nuanced impact of AI on creators. 

\bibliographystyle{ACM-Reference-Format}
\bibliography{ref}

\end{document}